\documentclass[aps,prl,reprint,amsmath,amssymb]{revtex4-2}
\usepackage{blindtext}

\usepackage[english]{babel}

\usepackage[letterpaper,top=2cm,bottom=2cm,left=3cm,right=3cm,marginparwidth=1.75cm]{geometry}

\usepackage{amsmath}
\usepackage{graphicx}
\usepackage[colorlinks=true, allcolors=blue]{hyperref}
\usepackage[font=small,labelfont=bf]{caption} % For better caption control
\usepackage{xcolor}

\usepackage{placeins}
\usepackage[utf8]{inputenc}  % For special characters
\usepackage[T1]{fontenc}     % Proper character output
\usepackage{subcaption}      % For subfigures
\usepackage{csquotes}        % Recommended when using babel or polyglossia
\usepackage{bm}              % Bold math

\usepackage{etoolbox} % For modifying bibliography entries
\usepackage{xpatch}   % For patching macros

\begin{document}

\preprint{APS/123-QED}

\title{Solving the inverse problem for the design of omnidirectional lenses with an ultimately high numerical aperture}

\author{Shay Rabani}
\affiliation{Faculty of Electrical Engineering, Tel Aviv University, Tel Aviv 69978, Israel.}

\author{Kirill Shvalb}
\affiliation{Faculty of Physics, Lomonosov Moscow State University, Moscow, 119991, Russia.}

\author{Boris Malomed}
\affiliation{Department of Physical Electronics, Faculty of Engineering, Tel Aviv University, Tel Aviv 69978, Israel; Instituto de Alta Investigaci\'{o}n, Universidad de Tarapac\'{a}, Casilla 7D, Arica, Chile.}

\author{Tal Carmon}
\affiliation{Department of Physical Electronics, Faculty of Engineering, Tel Aviv University, Tel Aviv 69978, Israel.}

% Date
\date{\today}  % This will display the current date

\begin{abstract}
We introduce a new type of lens that focuses a plane wave into a spherical one, where light comes from all directions. Our method also suggests the design of ideal optical tweezers or, in the reverse direction, photo-detection of nearly all of the light emitted from an omnidirectional source. Our design trades off simple isotropic non-magnetic materials for requiring a high refractive index. Despite many challenges, we believe that our proposed lens can evolve, with optimization and some help from conventional lenses, to fabricable lenses that will help in boosting collection efficiency of photons in ion-photon quantum experiments.
\end{abstract}

\maketitle

\section{Introduction}
For a couple of decades, there has existed a direct solution for engineering a spatial distribution of optical characteristics of the system that allows light rays to propagate along predetermined paths \cite{Pendry2006}. This is known as transformation optics, which makes use of vectorial anisotropic permittivity and permeability provided by meta-materials. Later, nonmagnetic cloaks \cite{Pendry2006, Jacob2008} and hyperlenses \cite{Jacob2007} were suggested, where the permeability is an isotropic scalar, but the permittivity is still vectorial, being supplied by dielectric meta-materials. Our approach differs from those techniques \cite{Pendry2006, Jacob2007}, as we rely on using solely ordinary materials with scalar isotropic permittivity, which is relatively easier to implement than the settings that incorporate meta-materials. In comparison to the design of optical “black holes” in Ref. \cite{Pendry2006, Narimanov2009}, our approach is different too. For example, an optical black hole \cite{Pendry2006, Narimanov2009} is based on a spherically symmetric structure, which implies that a unidirectional source at its center will emit light in all directions after passing through it. In contrast, our NA2 lens turns an omnidirectional emitter into a unidirectional one, where the upper limit for light collection by a photo-diode attains $100\%$. There are certain challenges in our current design, particularly the need to use a high refractive index \cite{Khurgin2022}. But these challenges might be mitigated with the help of regular lenses that prefocus the light before approaching our lens. Quantum experiments, such as those relying on trapped ions \cite{Leibfried2003}, improve with the percentage of light collected from an ion and transferred to a successive stage (e.g., a photodiode, another ion, or an optical fiber). This light collection efficiency can be improved by using high-numerical aperture lenses \cite{Monroe2014} or using an optical cavity \cite{Raimond2001, Rosenblum2017}. Despite many challenges, our long-term vision is that lens technology will evolve to collect light omnidirectionally, particularly in cases where light is emitted from an atom, an ion, or a quantum dot. An example related to ions is a quantum logic gate that utilizes the Hong-Ou-Mandel (HOM)  effect. This effect involves the quantum interference of two indistinguishable photons on a beam splitter, followed by their photo-detection. Assuming a photo-collection efficiency per photon of $50\%$, only one out of 4 measurements (on average) will successfully yield a reading from the two photons (i.e., $0.5^2 = 0.25$). A lens that collects nearly $100\%$ of the light can improve the upper success rate of such experiments to almost $100\%$. Furthermore, our inverse design method can be readily adapted to other settings that may be sought for in various optical devices, a relevant example being the focusing of a Gaussian beam (rather than the plane wave) into a perfect spherical wave and an omnidirectional invisibility cloaking where light bending is adiabatic and small index contrast might be within fabrication limits.” 

Ray-tracing methods, where ray propagation is calculated for a given refractive index distribution, have recently become common in commercial products [e.g., Zemax, Comsol, etc.]. In contrast, the inverse problem \cite{Deckelnick2011, Borghero2016, Pilozzi2018, Wang2023, Macdonald2023, Ito2024}, where a desired ray propagation scheme is provided to the designer, and the refractive index pattern required for maintaining the scheme should be produced (eventually, in the form of the program file for the 3D printer), has been rarely studied despite its potential to extend current technologies covering imaging and non-imaging optics, in fields ranging from optical interconnects to optical tweezers and concentrators for photovoltaic cells. For example, while current lenses collect or focus light from $180^o$, omnidirectional lenses, focusing light from $360^o$, would enable optical tweezers\cite{Nieminen2007, Moffitt2008, Polimeno2018, Pesce2020} that wrap up the particle with light in all directions and high-fidelity optical interrogation of a single-atom qubit \cite{Finkelstein2024}. Similarly, holographic tweezers can be improved for manipulating nanoparticles.

In addition to achieving a dry numerical aperture of 2 using a single lens, the omnidirectional lens would illuminate areas currently left dark on the tweezed particle, thus enhancing the sensitivity of the tweezers. Such tweezers are now regarded as the most sensitive method to measure small forces,
including forces applied by single molecules \cite{Zhang2013, AlBalushi2015, Koch2017, Zaltron2020, Favre2022, Yang2022}; among them, a DNA molecule \cite{McCauley2007, Aathavan2009, Capitanio2013, Alcon2024} and protein molecules \cite{Bustamante2020, Li2021, Sanchez2022}. 

3D printers were recently extended to work with glasses \cite{Marchelli2011, Schubert2014, Klein2015, Kotz2017, Zhang2020, Wen2021, Li2023}
and very recently to allow printing of spatially varying refractive indices \cite{Campbell2015, Zhang2016, Isakov2016, Isakov2018, Oh2020, DyllaSpears2020, Ketchum2022}. Unlike traditional subtractive manufacturing techniques, such as lens polishing, in the framework of which controlling the internal refractive index is a challenge, 3D printers belong to the family of additive manufacturing devices, in which samples are built layer by layer, allowing one to precisely control the structure of the internal refractive index. While it is straightforward to print standard components, such as waveguides, the transformative impact of printers lies in their ability to produce devices with arbitrary shapes that traditional manufacturing techniques cannot provide. Thus, our inverse design of the refraction index pattern, which is able to maintain a desirable ray family, comes at a timely moment, enabling not only explorative theoretical studies, but also printing practical appliances. Therefore, it seems that our designs can already address RF applications using current printing technology \cite{Yi2016, Isakov2018, Wu2018, Proyavin2024}
and with some optimization for low refractive index contrast, our method can provide designs for visible light using printing technologies \cite{DyllaSpears2020, Ketchum2022}.

The general difficulty in the solution of inverse problems is, roughly speaking, the same as the difficulty in the classical calculus, where performing integration is more challenging than the straightforward differentiation. Another common problem is that inverse problems tend to be ill-posed ones, i.e., the solution of such problems is often subject to an instability, while the solution of the respective direct problem may be quite stable \cite{Kabanikhin2008, Hofmann2013, Nashed2016}. As concerns the problem of the restoration of the distribution of the refractive index which gives rise to a predetermined family of rays, in the framework geometric optics, it was previously addressed, in an analytical form, only in some particular cases of isotropic two-dimensional patterns of the refractive index \cite{Borghero2005, Borghero2011}. Here, starting from particular analytical solutions, we develop generic three-dimensional numerical ones for the inverse problem, making them appropriate for securing any required ray design, much like versatile solvers available today do for the forward problems. Of course, problems with conditions that violate thermodynamic laws, such as the brightness law, remain unsolvable, with or without our solver.

Below, we define the problem, followed by the governing equations and their solution algorithm. The analysis aims to design a lens which is able to perform complete conversion of a spherical wave into a plane one and vice versa.

\section{The two-dimensional Eikonal Equations - the relationship between the refractive-index and ray-path patterns}

The starting point is the Eikonal Equation \cite{eikonal_equation}, which determines the shape of light rays in the plane of $(x,y)$, as determined by the refractive-index pattern $n(x, y)$. Here we use the Eikonal equation in the two-dimensional form:

\begin{equation}
\frac{d}{ds}\left( n \frac{d\textbf{r}}{ds} \right)=\nabla n,
\label{eq:eikonal}
\vspace{10pt}
\end{equation}
where $n(x,y)$ is the spatial pattern of the local refractive index, $\nabla n$ is its gradient, $\textbf{r}=(x, y)$ is the vector corresponding to a particular point belonging to the ray, and $s$ is the length of the arc along the ray. In the component form, vector Eq. (\ref{eq:eikonal}) can be replaced by a system of two scalar ones:

\begin{equation}
\frac{d}{ds}\left( n(x, y) \frac{dx}{ds} \right)=\frac{dn}{dx}
\label{eq:eikonal_x}
\end{equation}

\begin{equation}
\frac{d}{ds}\left( n(x, y) \frac{dy}{ds} \right)=\frac{dn}{dy}
\label{eq:eikonal_y}
\vspace{10pt}
\end{equation}

\begin{figure}[htbp]
    \begin{subfigure}[b]{0.45\textwidth}
        \includegraphics[width=8cm,height=6cm]{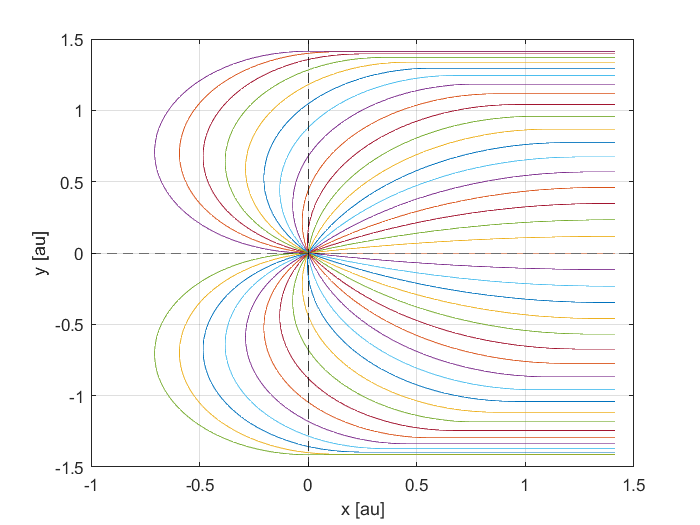}
        \caption{}
        \label{fig:DesireableFamilyRays_2D}
    \end{subfigure}
    \hfill
    \begin{subfigure}[b]{0.45\textwidth}
        \includegraphics[width=8cm,height=6cm]{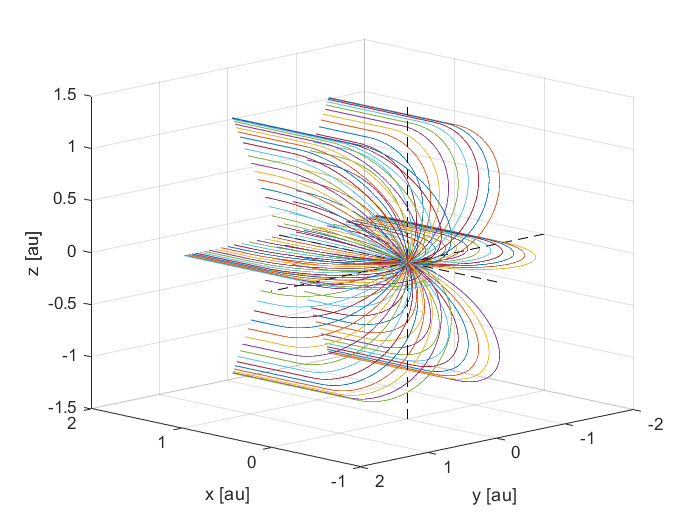}
        \caption{}
        \label{fig:DesireableFamilyRays_3D}
    \end{subfigure}
    \caption{\textbf{Desirable Family Rays to our design criteria for a dry NA2 lens}. (a) 2D ray tracing. (b) 3D ray tracing.}
    \label{fig:DesireableFamilyRays}
\end{figure}

The objective is to predict a pattern of $n(x,y)$ that will (approximately) transform a family of rays emitted from the origin, $(x, y)=(0, 0)$, in all directions, into a family of rays which asymptotically propagate parallel to the x axis. To this end, we attempt to solve the inverse problem for the Eikonal equation. We illustrate the target ray family by the plot shown in Fig. \ref{fig:DesireableFamilyRays}.

We require a family of rays, entering our lens as a plane wave, to focus into a spherical wave which is characterized by light radiating uniformly from all directions, encompassing a full solid angle of 4$\pi$ steradians, and focusing into a point. Similarly, our lens can operate in the opposite direction and map a point source to a plane wave.
The color lines in Fig. \ref{fig:DesireableFamilyRays} represent the rays which start at an arbitrary exit angle from the origin $(x,y,z)=(0,0,0)$ and eventually propagate parallel to the $x$ axis, while dashed black lines denote the $x$ and $y$ axes.

Consequently, our design includes an empty sphere at the focus. The dimensionless diameter of this sphere is $X$.

As one can see in Fig. \ref{fig:DesireableFamilyRays}, the rays farthest away from the optical axis (e.g., the ray that passes at $x=y=1.49$) are bent by $180^{\circ}$, redirecting them opposite to the direction of the incident beam. This specific case here where light makes a 'U-turn' uniquely distinguishes our NA2 lens from current lens technology where maximal ray tilt is limited to $90^{\circ}$ (instead of $180^{\circ}$ or higher as in our method).

\section{The geometric analysis}

\begin{figure*}[htbp]
\centering
\includegraphics[width=0.55\linewidth]{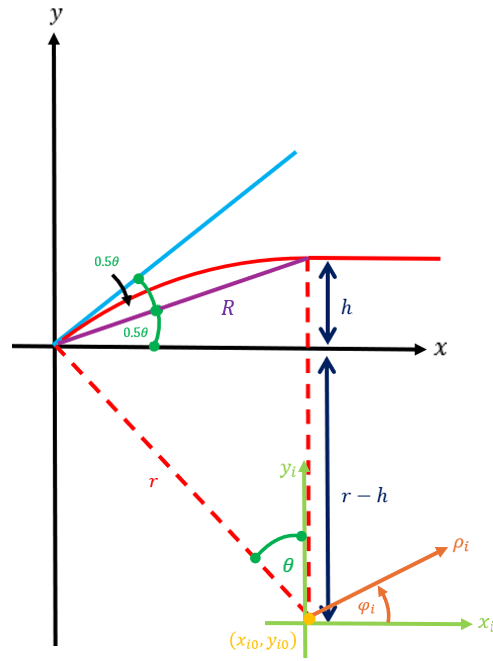}
\caption{\label{fig:FullAnalyticGeometryRayPath} \textbf{The geometric structure of the ray path} (see a detailed description in the text)}
\end{figure*}

As said above, each ray emitted from the origin, after passing the lens, must propagate parallel to the x-axis. Therefore, the ray path is divided into two segments: the first one is a slice of the circle's arc, which ends at the transition point that connects it to the second segment, running parallel to the $x$ axis. The structure of the compound ray is shown in Fig. \ref{fig:FullAnalyticGeometryRayPath}.

\vspace{10pt}
In Fig. \ref{fig:FullAnalyticGeometryRayPath}, the continuous red line represents the ray path, dashed red lines designate radii of the circle, the pale blue line is the tangent to the arc at the origin, the upper blue arrow designates height $h$ of the transition point above the $x$ axis, the lower blue arrow shows the vertical distance $r-h$ to the circle's center below the $x$ axis, and angle $\theta$ (made of $0.5\theta+0.5\theta$) determines the orientation of the ray at the origin. The angle between the radii is also equal to $\theta$, due to the fact that the sides of the two angles are mutually perpendicular. The black lines $(x,y)$ are the Cartesian coordinate system in which all the rays originate from the center, the pale green lines $(x_i,y_i)$ are the coordinate system with origin at $(x_{i0},y_{i0})$, i.e., at the center of the circular segment, and the orange lines $(\rho_i,\varphi_i)$ are the corresponding polar coordinate system. \\

Our aim here is to express the geometric parameters solely in terms of angle $\theta$.

\subsection{Creating a plane wave with uniform intensity}
The condition that we want to create a lens which not only focuses light over the panoramic view of $360^o$, but also turn input isotropic spherical wave into the output plane wave with uniform intensity, imposes the additional geometrical constraint. 

Using this constraint, we obtain one for $h$. Physically. it determines the transition point where the ray switches from the circular trajectory into the straight line.

In Fig. \ref{fig:FullAnalyticGeometryRayPath} the continuous purple segment $R$ designates the distance of the switching point from the origin.
The angle between this segment and the $x$ axis is $\theta/2$, due to the fact that the lower angle $\theta$ is an angle of the head in the isosceles triangle and its right side is perpendicular to the $x$ axis.

To obtain a relation between $R$ and $h$ we use the trigonometric relation in the upper triangle:

\begin{equation}
R=\frac{h}{\sin \left( \theta/2 \right)}\equiv h\sqrt{\frac{2}{1-\cos\theta}}.
\label{eq:h_and_R}
\vspace{10pt}
\end{equation}
%While we have used in equation (\ref{eq:h_and_R}) a double angle %identity.

As the intensity of the spherical wave decreases $\sim\frac{1}{R^2}$\cite{eikonal_equation} as it spreads out from the source, we have to omit the dependence of $R$ on $\theta$, with the aim of creating a plane wave with uniform intensity. Otherwise, different rays would carry different intensities, depending on the orientation of the ray at the origin. To meet this condition, we need to choose:

\begin{equation}
h=\sqrt{1-\cos\theta},
\label{eq:h}
\vspace{10pt}
\end{equation}
according to Eq. (\ref{eq:h_and_R}).\\

Note that the scaling invariance of the present setup implies that it can be enlarged or shrunk while preserving its shape.

\subsection{Analytic expressions for the ray paths}

We use the trigonometric relation in the lower triangle in Fig. \ref{fig:FullAnalyticGeometryRayPath} to get:

\begin{equation}
\cos\theta=\frac{r-h}{r}.
\label{eq:costheta}
\vspace{10pt}
\end{equation}
Then, the combination of Eqs. (\ref{eq:h}) and (\ref{eq:costheta}) yields an expression for $r$:

\begin{equation}
r=\frac{1}{\sqrt{1-\cos\theta}}.
\label{eq:r}
\vspace{10pt}
\end{equation}
Finally, we obtain coordinates $(x_0, y_0)$ of the circle's center:

\begin{equation}
\begin{aligned}
x_0=\frac{\sin\theta}{\sqrt{1-\cos\theta}} \\
y_0=-\frac{\cos\theta}{\sqrt{1-\cos\theta}}
\label{eq:centerpoint}
\vspace{10pt}
\end{aligned}
\end{equation}
Thus, an explicit analytical equations for the circular segment of the ray is written as

\begin{equation}
\begin{aligned}
x(s)=x_0+r\cos \left( \frac{s}{r} - \tilde{\theta} \right) \\
y(s)=y_0+r\sin \left( \frac{s}{r} - \tilde{\theta} \right)
\label{eq:circle}
\vspace{10pt}
\end{aligned}
\end{equation}
Here, as said above, $s$ is the length of the arc along the ray, and we define $\tilde{\theta} \equiv \theta + \frac{\pi}{2}$ to meet the initial condition, $x(s=0)=0$ and $y(s=0)=0$. 

\vspace{10pt}
The straight segment of the path, running parallel to the $x$ axis, is:

\begin{equation}
x>x_0, \hspace{1cm}
y=h=\sqrt{1-cos\theta}
\label{eq:paralleltox}
\vspace{10pt}
\end{equation}

\section{Calculating the refractive-index pattern}
As said above, to produce the ray paths of Eq. (\ref{eq:circle}), we have to design a lens with an appropriate spatial pattern of the refractive index. To this end, we need to solve Eqs. (\ref{eq:eikonal_x}) and (\ref{eq:eikonal_y}).
For this purpose, we use the coordinates defined in Fig. \ref{fig:FullAnalyticGeometryRayPath}, for a particular $i$-th ray trajectory. Note that centers $(x_{i0},y_{i0})$ are different for different rays, hence the coordinate systems $(x_i,y_i)$ and $(\rho_i,\varphi_i)$ are also different, being attached to particular rays.\\

The relation between the different coordinates is written as:

\begin{equation}
\begin{aligned}
x = x_{i0} + x_i = x_{i0} + \rho_i \cos\varphi_i \\
y = y_{i0} + y_i = y_{i0} + \rho_i \sin\varphi_i
\label{eq:coordinates_relations}
\vspace{10pt}
\end{aligned}
\end{equation}

For the circular segment of the trajectory, its length is
$s_i=\rho_i\varphi_i$, hence the derivative with respect to $\varphi_i$ is written as

\begin{equation}
\frac{d}{d s_i} = \frac{1}{r_i} \frac{d}{d\varphi_i}. 
\label{eq:d_ds}
\vspace{10pt}
\end{equation}

Next, we transform the underlying equations (\ref{eq:eikonal_x}) and (\ref{eq:eikonal_y}) in the coordinate system attached to the particular ray path, the result being

\begin{equation}
\frac{1}{r_i} \frac{dn}{d\varphi_i}(-\sin\varphi_i) + n\frac{(-\cos\varphi_i)}{r_i} = \frac{dn}{dx_i}
\label{eq:eikonal_x_i}
\end{equation}
\begin{equation}
\frac{1}{r_i} \frac{dn}{d\varphi_i}(\cos\varphi_i) + n\frac{(-\sin\varphi_i)}{r_i} = \frac{dn}{dy_i}
\label{eq:eikonal_y_i}
\vspace{10pt}
\end{equation}

Further, the linear combination of the above equation, taken as
$(\ref{eq:eikonal_x_i})\times\cos\varphi_i + (\ref{eq:eikonal_y_i})\times\sin\varphi_i$, we obtain the simple equation,

\begin{equation}
\frac{dn}{d\rho_i} = -\frac{n}{r_i}.
\label{eq:dn_drho}
\vspace{10pt}
\end{equation}
To derive it, we have used the chain rule:

\begin{equation}
\begin{aligned}
\frac{dn}{d\rho_i} = \frac{dn}{dx_i} \frac{dx_i}{d\rho_i} + \frac{dn}{dy_i} \frac{dy_i}{d\rho_i} \\
\frac{dx_i}{d\rho_i} = \cos\varphi_i,  \hspace{0.5cm}
\frac{dy_i}{d\rho_i} = \sin\varphi_i
\label{eq:dn_drho_chain_rule}
\vspace{10pt}
\end{aligned}
\end{equation}
Thus, we have reduced the system of coupled PDEs (\ref{eq:eikonal_x_i}) and (\ref{eq:eikonal_y_i}) into the single equation (\ref{eq:dn_drho}) with the single variable $\rho_i$. Then, to produce the required refractive-index pattern, it is straightforward to apply the gradient-descent (GD) optimization method \cite{Bauer1997, Figueiredo2007, Li2012, Mu2023}, with boundary condition along the ray-path of $\theta=0^o$ to Eq. (\ref{eq:dn_drho}).

While the GD method is more commonly used in the context of machine learning and optimization problems, it can be adapted to solve certain types of differential equation by minimizing the error of approximate solutions.
To apply it in the present context, we define a loss function based on the difference between the calculated gradient $\frac{dn}{d\rho_i}$ and the analytical expression $-\frac{n}{r_i}$ given by Eq. (\ref{eq:dn_drho}).Therefore, the loss function (error function) is:

\begin{equation}
L(n) = \frac{dn}{d\rho_i} + \frac{n}{r_i}.
\label{eq:loss_equation}
\vspace{10pt}
\end{equation}
This error function was set such that the error on the refractive index is less than $0.0029$. This value was chosen because it provides an error much smaller than the figure resolution within a few seconds of the simulations. Upon need, we can achieve better accuracy at the cost of a few minutes of run time for the simulation.

\begin{figure}[ht]
\centering
\includegraphics[width=8cm,height=6.5cm]{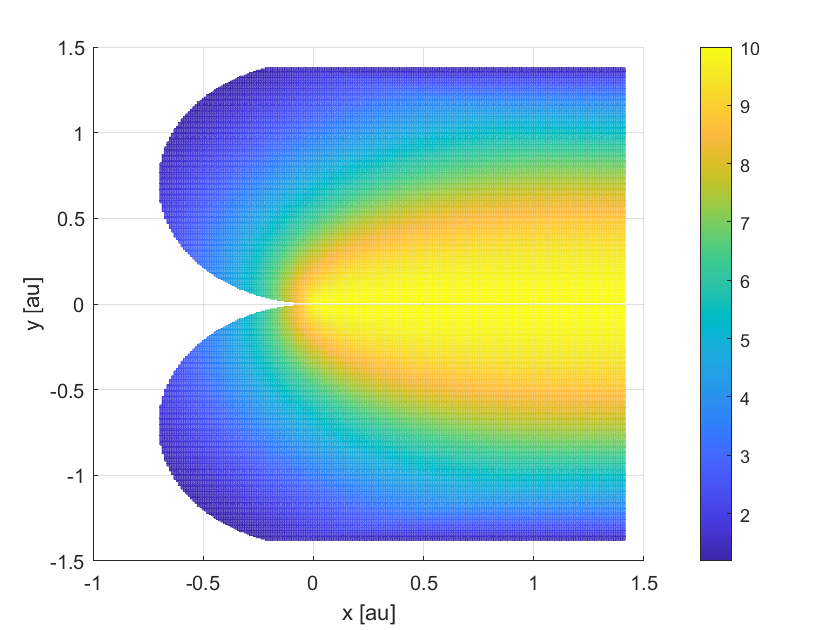}
\caption{\label{fig:RefractiveIndexPatternGridpng} \textbf{Dry NA2 lens design.} Refractive-index distribution solution for a perfect NA2 lens.}
\end{figure}

We choose the refractive index $n=10$ at $x>0, y=0$ to prevent a situation in which the required refractive index elsewhere will drop below 1. The latter condition implies that our setup can be built without the use of meta-materials, thus making the implementation relatively easier. 
Using this boundary condition, we produce the mesh pattern shown in Fig. \ref{fig:RefractiveIndexPatternGridpng}.
Note that the minimum value of the refractive index (dark blue) is $n=1.186$.

Our approach includes high values, $n=10$, of the refractive index on the optical axis of the lens, which might be first realized in radio frequency [RF] bands, where high refraction was recently reported \cite{Choi2011}. As for implementing our method in the IR and visible parts of the spectrum, where the use of high refractive indices is challenging \cite{Khurgin2022}, we are working on mitigating the high-index requirements by using a combination of regular lenses, with our lens.

Another challenge is having a high refractive index at the boundary between air and the lens. In such high index-contrast transitions, a considerable portion of the light may be reflected, rather than transmitted. As is known, anti-reflective coating is used to mitigate such unwanted reflections. In this connection, we note that anti-reflection coating for lenses based on germanium, whose refractive index is 4, is currently common in many products.

While Fig. \ref{fig:RefractiveIndexPatternGridpng} presents the resulting refractive-index pattern in the 2D form, its 3D counterpart is obtained by rotating Fig. \ref{fig:RefractiveIndexPatternGridpng} around the horizontal symmetry axis.

\section{Conclusion}
Our numerical solution offers the new technique, in the framework of which the optical design software includes not only trace rays for a given refraction-index pattern but solves the inverse problem to provide the index pattern maintaining the desirable global ray family. We are currently continuing this work toward printing RF lenses with a numerical aperture equal to 2, which allows the transformation of a collimated beam into a converging spherical wave that can reach a detector from all directions. Our work, together with recent developments in 3D printing technology, should open a way to the fabrication of a new type of optical components in which the local refraction can vary arbitrarily in 3D, allowing applications which are not supported by the current fabrication techniques.

\section*{Acknowledgments}
We thank Barak Dayan, Hagai Eisenberg, Ady Arie, Roee Ozeri and Ofer Kfir for the fruitful discussion.
We  acknowledge valuable discussions with Prof. Ady Arie and the help of Shai Zucker in numerical calculations. The work of B.A.M. was supported, in part, by Israel Science Foundation through grant No. 1695/22. The work of T.C. was supported through the United States–Israel Binational Science Foundation (NSF-BSF) through grant No. 2020683 and the Israeli Science Foundation through grant No. 537/20.

\bibliography{sample} % Tell bibtex which .bib file to use (this one is some example file in TexLive's file tree)

\end{document}